\documentclass[aps,prl,twocolumn,superscriptaddress]{revtex4-1}
\usepackage{amssymb}
\usepackage{bm}

\usepackage[english]{babel}
\usepackage{amsmath,amsfonts,graphicx,epsfig}

\begin{document}

\title{Microwave control of coupling parameters in spinor alkali condensates}
\author{V. Cheianov}
\affiliation{Instituut-Lorentz, Universiteit Leiden, P.O. Box 9506, 2300 RA Leiden, The Netherlands}
\author{A. L. Chudnovskiy}
\affiliation{1. Institut f\"ur Theoretische Physik, Universit\"at Hamburg,
Jungiusstr 9, D-20355 Hamburg, Germany}

\date{\today}

\begin{abstract}
We propose a protocol which utilises radio frequency magnetic pulses in order to
tune the effective two-particle scattering amplitudes for alkali atoms in the $F=1$
hyperfine ground state. Unlike the Feshbach resonance method, the proposed 
protocol preserves with  controllable accuracy the global rotational symmetry 
in the spin space offering access to a broad region 
of the phase diagram of the rotationally-symmetric spinor Bose condensate.
Examples of $^{41}$K and $^{7}$Li are considered and it is shown that for these
atoms sufficient variation in the effective coupling constants 
can be achieved in order to explore phase transitions between different 
symmetry-broken phases of the condensate.  
\end{abstract}

\maketitle

Utracold atomic systems serve as potent quantum laboratories, where an amazing 
range of many-body hamiltonians can be realised and investigated 
owing to the precise spacial and temporal control over the interactions.   
Of particular interest is the possibility to fabricate bespoke strongly interacting many 
body systems with a high spin symmetry, thus paving a way to the investigation of 
mechanisms of spontaneous symmetry breaking, symmetry changing quantum phase transitions, 
spin liquid states and other topological states of matter  
\cite{Stamper-KurnReview2013,Lewenstein07,Hermele09,Hermele10,Hermele16}. In 
order to use the full potential offered by cold atoms it is desirable to have 
control over effective interaction strength while preserving the high degree of 
symmetry of the interaction. The most popular way of tuning the interaction, the 
Feshbach resonance, does not offer such a possibility because of the symmetry 
breaking Zeeman splitting \cite{Pethik-Smith}.  In some cases, the optical Feshbach 
resonance can be used as an alternative way of control over interactions  
\cite{Fedichev,Bohn97,Bohn99}, which has been  demonstrated for $^{87}$Rb, $^{172}$Yb, 
and $^{88}$Sr  atoms \cite{Theis04,Enomoto08,Nicholson15,Sandor16}. 

Here we investigate an alternative way of symmetry-preserving control over the 
coupling constants utilising radio-frequency (RF) dressing technique. Our interest is 
inspired by the successful use of the radio-frequency radiation for the manipulation of 
the hyper-fine Zeeman split levels of the atomic ground state manifold. This technique 
enabled, in particular, creation of topologically nontrivial structures in the magnetic 
BEC condensate \cite{WilliamsNature,MattewsPRL,AndersonPRL}. It was also used for the
study the nonequilibrium population dynamics of different spin sates 
\cite{Liu09_1,Liu09_2,Zhao14,Bookjans11}.
We focus on the spin-1 alkali condensate as a minimal nontrivial example with 
competing spin-dependent interactions. Spin-symmetric interactions are 
characterized by two interaction constants, which in turn are determined by the 
spin-scattering length in the total momentum $F=0$ and $F=2$ channels.  
Different quantum phases are realized depending on the relation between the two 
coupling constants  \cite{Ho98}. Therefore, experimental exploration of the 
quantum phase diagram requires their tunability.  
In this work, we find that a train of radio-frequency pulses can be used to 
control {\em inter-atomic} interactions giving access to a broad range of 
coupling constants. Moreover, we design a protocol of the time-dependent 
polarization of RF field that suppresses the quadratic Zeeman coupling, thus 
leaving the spin-rotational symmetry intact. 
Extended to higher spin, our approach may prove useful for the implementation of 
tunable $SU(N)$ symmetric interactions that are necessary for the experimental 
creation of exotic topologically nontrivial spin liquid phases 
\cite{Hermele09,Hermele10,Hermele16}. 

We begin by considering the basic microscopic process of low-energy collision of 
two identical alkali atoms. The direct interaction between nuclear spins in such 
a process can be neglected \cite{Pethik-Smith}, therefore  the outcome of the 
collision is determined by scattering lengths in the singlet and triplet 
electron spin channels,  $a_1$ and $a_3$. Assuming that each atom in the wave 
zone is in its hyperfine ground state, the inter-atomic collision is 
characterized by the set of scattering lengths corresponding to the channels of 
different total angular momentum $F$.   The scattering problem is encoded in the 
radial $l=0$ center-of-mass equation 
\begin{eqnarray}
\nonumber && 
 \left[-\frac{1}{2\mu} \frac{d^2}{dr^2} + {^1V}(r)\, {^1}\hat {P} + {^3V}(r)\,  ^3\hat {P} + \right. \\ 
 && \left.  
 A ( \hat {\mathbf I}_1\cdot \hat { \mathbf S}_1+\hat {\mathbf I}_2\cdot \hat {\mathbf S} _2)\right] \psi 
= E\psi
\label{two-atom-scatt}
\end{eqnarray}
Here $\mathbf S_i$ is the spin of the $i$th particle, $\mathbf I_i$ is its 
nuclear spin and ${^m\hat P}$ is the 
projector on the two-atom spin state of multiplicity $m$, and $\mu$ denotes the 
reduced mass of the two atoms.  At low energies, ${^m V(r)}$ can be replaced by 
the zero-range pseudopotentials that impose 
the appropriate scattering length in the corresponding channel through the 
Bethe-Peierls boundary condition \cite{Petrov}. 
The constant $A$ is the hyperfine constant of an individual atom.
The two-atom wave function $\psi$ is a reducible spinor of size $ 4\times 
(2I+1)^2.$ 
Here we are interested in the spin-1 bosons, which are realized by alcali atoms 
with the nuclear spin $I=3/2$.

For given kinetic energy $E$ the wave-zone motion can be characterized by a set 
of four good quantum numbers  $F_1,F_2, F, M_F$ that is the total momentum of 
each atom, the total momentum of the pair and its projection.
Introducing the projector $\hat P_{F_1F_2}^F$ onto the subspace of given 
values of momenta we write the wave zone dispersion relation 
\begin{equation}
 \sum_{F_1=1}^2\sum_{F_2=1}^2 \sum_{F=|F_1-F_2|}^{F_1+F_2}\left(\frac{k^2}{2m} + E_{F_1F_2} - E\right) \hat P_{F_1F_2}^F \psi = 0,
\end{equation}
where 
\begin{equation}
 E_{F_1F_2}= \frac{A }{2} \left[F_1(F_1+1) + F_2(F_2+1)\right] - 
 2 A,  
\end{equation}
and we chose the energy offset such that $E_{11} = 0.$
We are interested in the low energy  limit 
$ E\ll A$, 
where both atoms are in the lowest hyperfine state 
with $F_1=F_2=1$. In this limit the higher energy spin configurations can only contribute to 
scattering as evanescent modes. We introduce the corresponding values of the 
momentum 
\begin{equation}
 k = \sqrt{2 \mu E}, \qquad \kappa_{F_1F_2} = \sqrt{2\mu\left(E_{F_1F_2}-E \right)}.
\end{equation}
Furthermore, we  define the following matrix
\begin{equation}
\hat T(r) = (e^{-i k r} - S e^{i k r}) {\hat P}^0_{11}   +  \alpha 
\hat{P}_{22}^0 e^{-\kappa_{22} r}
\end{equation}
for the $F=0$ scattering channel and and seek a solution to Schroedinger's equation in the form
\begin{equation}
 \Psi(r) = T(r) v
\end{equation}
where $v$ is some arbitrary spinor with $F_z=0.$

For the $F=2$ channel we write
\begin{eqnarray}
\nonumber && 
\hat T(r) = (e^{-i k r} - S e^{i k r}) {\hat P}^2_{11}   +  
 \left(\alpha \hat{P}_{12}^2 + \beta {\hat P}_{21}^2\right) e^{-\kappa_{12} r} \\ 
 &&  + \gamma {\hat P}_{22}^2 e^{-\kappa_{22} r}
\end{eqnarray}
and assume that $v$ is any vector with $F_z=-2.$ The Bethe-Peierls boundary conditions read
\begin{equation}
^m{\hat P}\, \left( \Psi'(0) + \frac{1}{a_m} \Psi(0)\right) = 0, \qquad m=1,3
\end{equation}
and are to be used for the determination of the unknown coefficients $S,\alpha, \beta, \gamma.$ 
Since $S=e^{2 i\delta}$ the scattering length is found from 
\begin{equation}
 i k \frac{S+1}{S-1} = -\frac{1}{a} + o(k), \qquad k\rightarrow 0
\end{equation}


The result of the calculation is as follows.
In the $F=0$ channel
\begin{equation}
\frac{1}{a_{F=0}}= \frac{10 a_1+6 a_3-8 \lambda }{16 a_3 a_1-3 a_1 \lambda -5 a_3 \lambda }.
\label{a0}
\end{equation}
In the $F=2$ channel 
\begin{eqnarray}
\nonumber && 
\frac{1}{a_{F=2}}=\left[6\sqrt{2}a_3^2+26\sqrt{2}a_1a_3+16\lambda^2-(6\sqrt{2}+14)a_1\lambda-\right. \\ 
\nonumber && 
\left.
(10\sqrt{2}+18)a_3\lambda \right]\bigg/ \left[3a_1\lambda^2+13 a_3\lambda^2-(7\sqrt{2}+12)a_3^2\lambda-
\right. \\ 
&&  \left.  
(9\sqrt{2}+20)a_1a_3\lambda+32\sqrt{2}a_1a_3^2
\right], 
\label{a2}
\end{eqnarray}
where
\begin{equation}
 \lambda^2=\frac{\hbar^2}{2\mu A}.
 \label{lambda}
\end{equation}
One can see from Eqs. (\ref{a0}), (\ref{a2}) that the numerator vanishes at certain values of $\lambda$, signaling the resonantly diverging scattering length.  The scattering length varies strongly close to the resonance, and it changes sign on different sides of the resonance. In the rest of the paper we show how one can use rf-pulses in order to vary the effective hyperfine interaction constant $A$, thus changing the value of $\lambda$ by virtue of Eq. (\ref{lambda}).  We propose a protocol which enables an order of magnitude variation of the strength of effective hyperfine interaction while maintaining  the spin-rotational symmetry. 

The proposed protocol consists of periodic sequence of pulses with magnetic field within one period $T$ given by 
\begin{equation}
{\bf h}(t)=\sum_{n=1}^4 {\bf e}_n\eta(t-n  T/4). 
\end{equation}
Here the four unit vectors ${\bf e}_n$ form the vertices of a perfect tetrahedron, and the function $\eta(t)$ is defined by 
\begin{equation}
\eta(t)=\left\{\begin{array}{rll}
h_0 & \mbox{for} & 0\leq t < T/16, \\ 
-h_0 & \mbox{for} & 3 T/16\leq t <  T/4,\\ 
0 &  \mbox{otherwise}. &
\end{array}
\right. 
\label{eta_t}
\end{equation}
The rf-field introduces the periodic time dependent perturbation to the Hamiltonian 
$
H_{\mathrm{rf}}=g\mu_{\mathrm{B}}{\bf h}(t) \cdot {\hat{\bf S}}. 
$

The effective Hamiltonian is calculated as the {\em time-independent} Hamiltonian that results in the equivalent quantum evolution operator over the period of the sequence \cite{Fishman,Dalibard}
\begin{equation}
\hat{U}(T,0)=\prod_{n=1}^4 \hat{U}_n=e^{-i\hat{H}_{\mathrm{eff}}T},
\label{U}
\end{equation}
where 
\begin{eqnarray}
\nonumber && 
\hat{U}_n=\exp\left[i \frac{T}{16}(\Omega_R{\bf e}_n - A \, \hat{\bf I})\cdot\hat{\bf S}\right]\exp\left[-i  A \, \hat{\bf I}\cdot\hat{\bf S}\frac{T}{8}\right]\times \\  
&& 
\exp\left[-i\frac{T}{16} (\Omega_R {\bf e}_n + A \, \hat{\bf I})\cdot\hat{\bf S}\right], 
\label{Un}
\end{eqnarray}
where $\Omega_R=g\mu_{\mathrm{B}}h_0$ denotes the Rabi frequency in the magnetic field of the pulse. 
In accord with Eq. (\ref{U}), the effective energies are calculated from the eigenvalues $u_k$ of the evolution operator by 
\begin{equation}
E_k=(i/ T) \log u_k.
\label{eigenEnergies}
\end{equation}
The effective hyperfine splitting is defined as the difference between the 
largest and the smallest eigenenergies, 
$A_{\mathrm{eff}}=E_{\mathrm{max}}- E_{\mathrm{min}}$. The proposed protocol 
introduces a small quadratic Zeeman coupling, leading to a splitting of the 
energy levels with $M=0$ and $M=\pm 1$. The induced quadratic Zeeman splitting 
is determined by the difference of the two neighboring eigenvalues 
$E_{1,0}-E_{1,1}$ in the multiplet with the total angular momentum $F=1$.   
\begin{figure}
\hskip -0.1cm \includegraphics[width=.9 \linewidth]{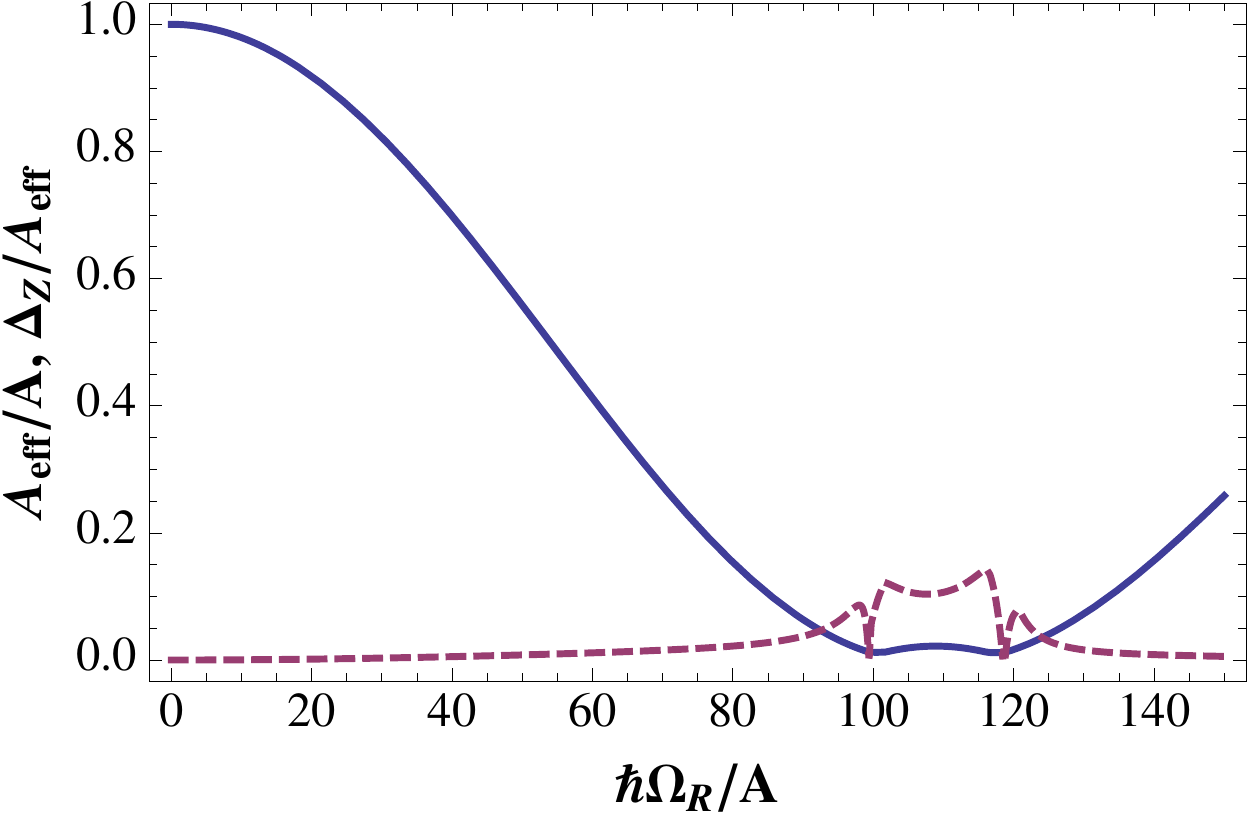}
\caption{(Color online).  Effective hyperfine coupling $A_{\mathrm{eff}}/A$ 
(solid line), and the quadratic Zeeman splitting (dashed line) as function of 
the Rabi frequency 
$\Omega_R$.   $A$ denotes the hyperfine coupling without external rf-driving. 
The calculations are performed for the cycle with the  period $T=\pi\hbar/A$ 
(dimensionless period $\tau=1/2$).}
\label{Fig_Aeff_hpulses}
\end{figure}

\begin{figure}
\includegraphics[width=0.9 \linewidth]{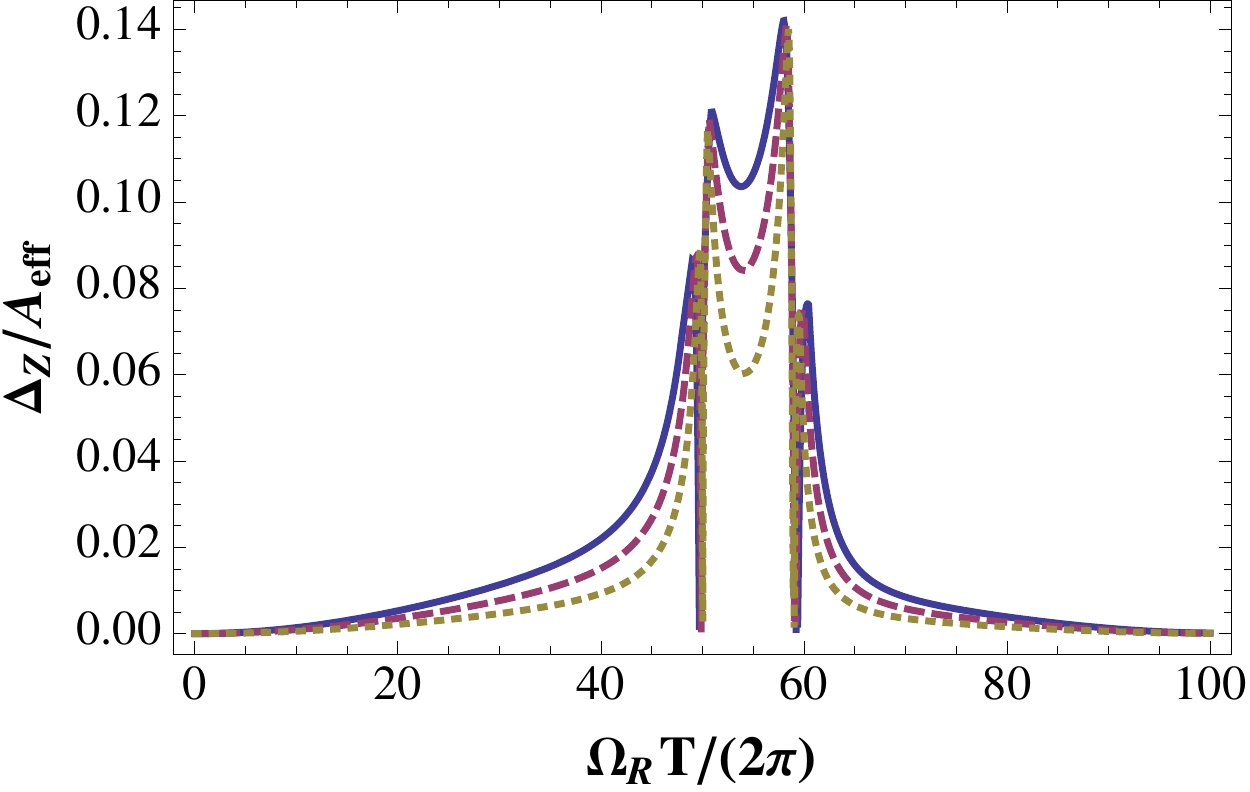}
\caption{(Color online).  Quadratic Zeeman splitting as function of the Rabi 
frequency 
$\Omega_R T$ for different dimensionless periods $\tau=1/2$ (solid line), 
$\tau=1/3$ (dashed line), $\tau=1/5$ (dotted line).    }
\label{Fig_Zeeman_Rabi}
\end{figure}

The effective Hamiltonian defined by Eqs. (\ref{U}), (\ref{Un}) is completely 
determined by two dimensionless parameters, which can be chosen as the 
dimensionless period of the pulse sequence $\tau= T A/(2\pi\hbar)$, and the 
dimensionless Rabi frequency $\hbar\Omega_R/A$.  
The change of the effective hyperfine coupling induced by the dressing with 
rf-pulses is shown in Fig. \ref{Fig_Aeff_hpulses}. 
One can see the suppression of the hyperfine coupling by an order of magnitude 
(solid line in Fig. \ref{Fig_Aeff_hpulses}). Especially interesting for 
practical applications are the points $\hbar\Omega_R/A\approx 16$, 
$\hbar\Omega_R/A\approx 21$, at which the quadratic Zeeman coupling is 
completely suppressed, hence the full spin rotational symmetry remains intact. 
The data in Fig. \ref{Fig_Aeff_hpulses} were calculated for the dimensionless 
period of the pulse sequence $\tau=0.5$. Changing the value of 
$\tau$ leads to the rescaling of the curves without changing their qualitative 
behavior. The quadratic Zeeman splitting for different values of $\tau$ is shown 
in Fig. \ref{Fig_Zeeman_Rabi}. 

\begin{figure}
\includegraphics[width=0.8 \linewidth]{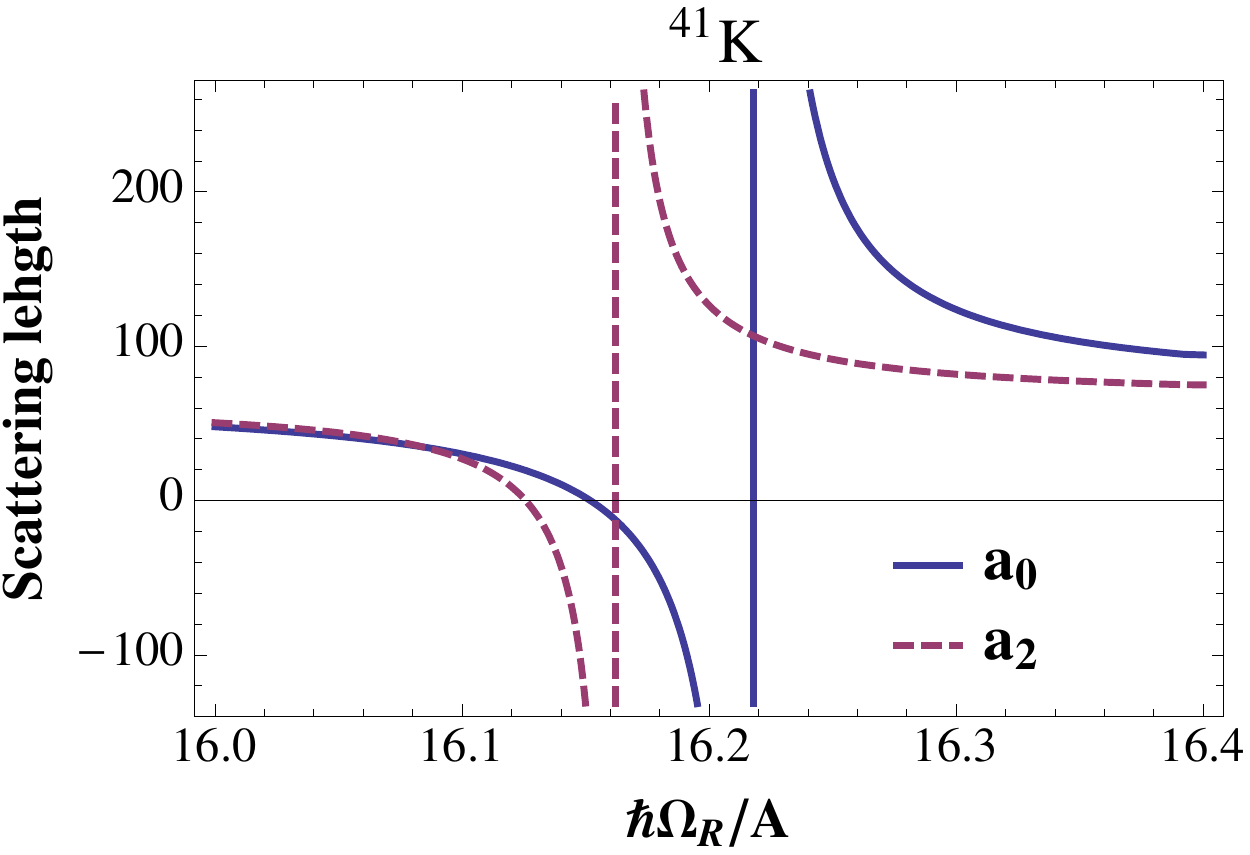}
\vskip 0.5 cm
\includegraphics[width=0.8 \linewidth]{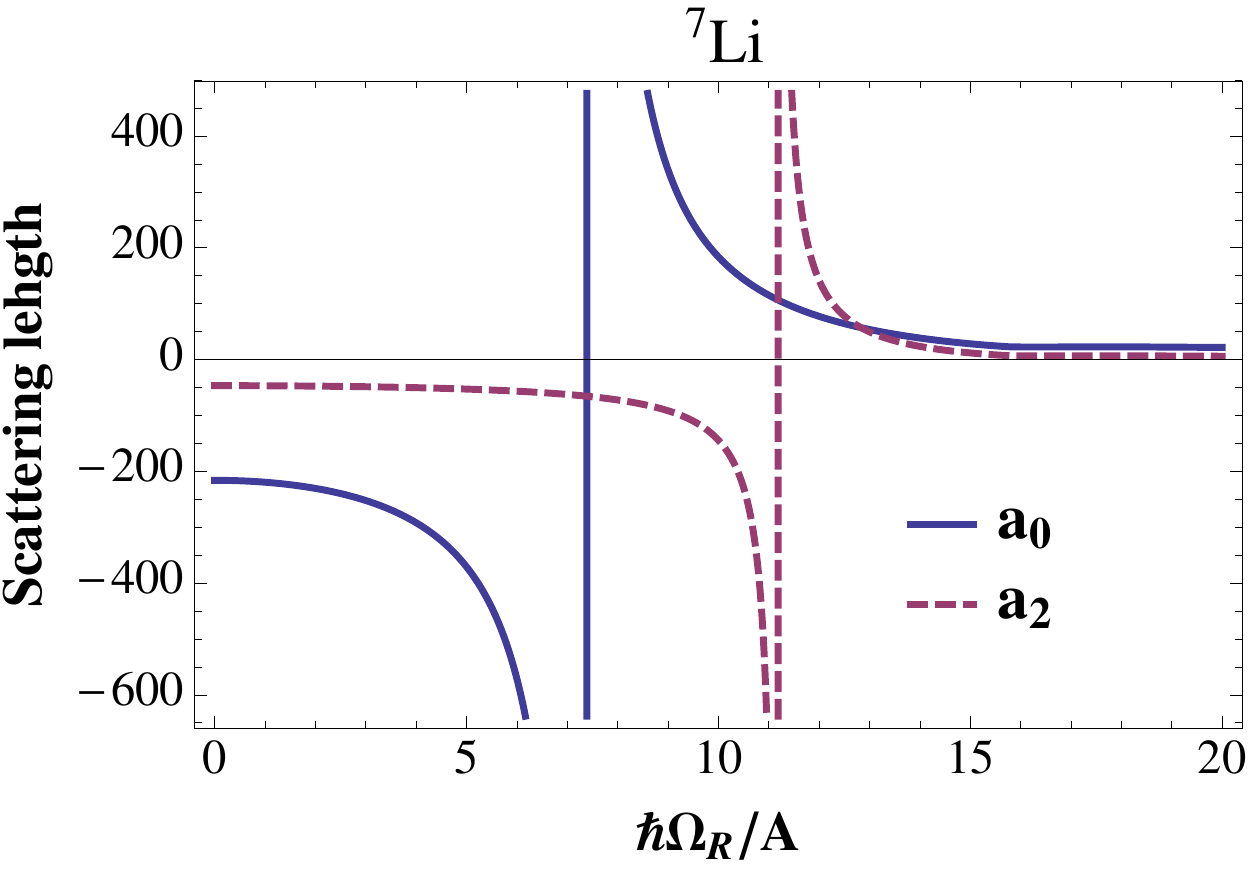}
\caption{(Color online). Upper panel: Two-particle scattering length for $F=0$ 
($a_0$, solid line), and $F=2$ ($a_2$, dashed line) channels as a function of 
the Rabi frequency $h$ in a single pulse (see the description of the rf-field 
protocol in the text)  calculated for spin-1 $^{41}$K atoms. In the region right 
to the resonance, one observes $a_0<0$ and $a_2>0$, hence the attraction in the 
$F=0$ channel is complemented by the repulsion in the $F=2$ channel. The 
parameters of $^{41}$K atoms  are taken as follows (in atomic units): $a_1 = 
85.5$, $a_3 = 60.5$, hyperfine coupling constant $A=2\pi\hbar \cdot (127  
\mathrm{MHz})$. The period of the pulse sequence is $T= 3\hbar/A$ corresponds to 
the frequency $\sim 266$MHz. The Rabi frequency close to the resonance 
$\Omega_R\approx 12.9$GHz, which corresponds to the magnetic field $\sim 
0.073$T. \newline
Lower panel: Two-particle scattering length for $F=0$ ($a_0$, solid line), and 
$F=2$ ($a_2$, dashed line) channels as a function of the Rabi frequency $h$ in a 
single pulse (see the description of the rf-field protocol in the text)  
calculated for spin-1 $^{7}$Li atoms. In the region between the two resonances, 
one observes $a_0>0$ and $a_2>0$, which indicates the region of a possible 
stable BEC condensate of Li-atoms. The parameters of $^{7}$Li atoms  are taken 
as follows (in atomic units): $a_1 = 35.5$, $a_3 = -17.0$, hyperfine coupling 
constant $A_0=2\pi\hbar \cdot (803.5 \mathrm{MHz})$. The duration of the single 
pulse $\tau= 3\hbar/(16 A)$ corresponds to the frequency $\sim 26.911$GHz. 
The Rabi frequency close to the resonance is $\Omega_R\approx 50.46$GHz, which 
corresponds to the magnetic field $\sim 0.29$T. }
\label{Fig_K41-Li7}
\end{figure}

It is important to note that the proposed protocol does not change the inter-atomic two-body interaction 
potential in the leading order of the high-frequency expansion in $1/\omega$, 
where $\omega$ is the fundamental frequency of the driving (in the considered 
case 
$\omega=2\pi/T$) \cite{Fishman,Dalibard}. Indeed, the Hamiltonian of two atoms 
in the driving field can be represented as 
$H=H_1+H_2+H_{\mathrm{int}}+H_{\mathrm{ext}}(t)$. In the proposed protocol, the 
interaction with external RF field for two atoms is written as  
$H_{\mathrm{ext}}(t)={\bf h}_{i}(t)(\hat{S}_{1i}+\hat{S}_{2i})$, hence it 
commutes with the interaction part of the Hamiltonian $H_{\mathrm{int}}$ (see 
Eq. (\ref{two-atom-scatt})). Performing the canonical transformation 
$\tilde{H}=U H U^{-1}$ with $U=\left(T e^{i\int H_{\mathrm{ext}}(t) dt}\right)$ 
we obtain $\tilde{H}= \tilde{H}_1(t)+\tilde{H}_2(t)+H_{\mathrm{int}}$. The 
interaction term remains unchanged, since it commutes with 
$H_{\mathrm{ext}}(t)$. In the proposed driving protocol, the transformation 
operator $U$ is periodic in time, which results in the periodic time dependence 
of the effective one-atom Hamiltonians $\tilde{H}_1$, $\tilde{H}_2$ with the 
frequency $\omega$. Therefore, the effective one-particle Hamiltonians  
$\tilde{H}_1$, $\tilde{H}_2$ allow the Fourier expansion 
$\tilde{H}_i(t)=\bar{H}_i+\sum_{n=1}^{\infty}\left(V_{i,n}e^{i n\omega t}+ 
V_{i,-n}e^{-i n\omega t}\right)$.  For $\hbar\omega$ larger than the hyperfine 
splitting,  the effective time-independent Hamiltonian can be constructed using 
the $(1/\omega)$ expansion as described in details in Refs. 
\cite{Fishman,Dalibard}. Thereby, the zero order term reads 
$\bar{H}_1+\bar{H}_2+H_{\mathrm{int}}$, where $\bar{H}_i$ denote the single atom 
Hamiltonian with renormalized hyperfine coupling. The corrections to the 
interaction part of the Hamiltonian appear only in the order of 
$1/\omega^2$ (see Eq. (C10) in Ref.\cite{Dalibard}).

To illustrate the proposed method of control over the inter-atomic scattering 
length, we calculated the change of the scattering length for two types of 
spin-1 atoms, that are often used for creation of the BEC. The spin-dependent 
scattering lengths for $^{41}$K atoms are shown in upper panel of  Fig. 
\ref{Fig_K41-Li7}. Particularly interesting is the region between the two 
resonances, where the negative scattering length in the $F=0$ channel is 
complemented by the positive scattering length in the $F=2$ channel. One can 
expect appearance of unconventional BEC in this regime. The lower panel in Fig. 
\ref{Fig_K41-Li7} shows the scattering lengths for ${^7}$Li atoms. The BEC of  
${^7}$Li is known to be unstable against collapse under usual conditions, 
because of the negative scattering lengths in both $F=0$ and $F=2$ channels. One 
can see in Fig. \ref{Fig_K41-Li7} (lower panel), that the application of the 
external RF field can drive  ${^7}$Li atoms in the region between the two 
resonances, where both scattering length become positive. One can expect a 
stable BEC of ${^7}$Li in that regime.  In both considered cases, the necessary 
RF frequencies and the Rabi frequencies associated with the magnetic field pulse 
lie within several GHz. Such conditions can be realized in the modern RF pulse 
generators. 

In conclusion, we proposed a method of control over the two-particle scattering 
length for spinful atoms, that ({\bf i}) allows large relative change of 
scattering length in the channels with different total angular momenta, 
including the change of the sign of the scattering length, ({\bf ii}) preserves 
the spin-rotational symmetry. Instead of the widely used magnetic and optical 
Feshbach resonances between the two-particle states in open and closed 
scattering channels, the proposed protocol is based on the change of the 
effective hyperfine coupling of a {\em single} atom dressed by the external 
RF field. Explicit calculations for $^{41}$K atoms show, that one can reach the 
regime, where the scattering lengths in $F=0$ and $F=2$ channels have different 
signs, which would allow experimental investigation of spinor Bose condensates 
with exotic magnetic orderings. Calculations for $^7$Li atoms demonstrate the 
regime, where both scattering length can change sign from initially negative 
without the external radiation to positive in the field, which provides a 
mechanism for the stabilization of spinor condensate of $^7$Li atoms against 
collapse.  Our findings may prove useful for experimental creation and 
investigation of spinor Bose condensates that could not be realized using 
magnetic or optical Feshbach resonances.

\end{document}